\documentclass[%
 aip,
 amsmath,amssymb,
 reprint,%
]{revtex4-1}

\usepackage{graphicx}% Include figure files
\usepackage{dcolumn}% Align table columns on decimal point
\usepackage{bm}% bold math
\usepackage[utf8]{inputenc}
\usepackage[T1]{fontenc}
\usepackage{mathptmx}
\usepackage{etoolbox}
\usepackage{siunitx}
\usepackage{booktabs}
\usepackage{multirow}
\usepackage[capitalise]{cleveref}
\usepackage[caption=false]{subfig}

\newcommand{\Rey}{Re}

\makeatletter
\def\@email#1#2{%
 \endgroup
 \patchcmd{\titleblock@produce}
  {\frontmatter@RRAPformat}
  {\frontmatter@RRAPformat{\produce@RRAP{*#1\href{mailto:#2}{#2}}}\frontmatter@RRAPformat}
  {}{}
}%
\makeatother
\begin{document}

\preprint{AIP/123-QED}

\title[Flow Development in Slender Converging Pipes]{Flow Development in the Entrance Region of Slender Converging Pipes}
% Force line breaks with \\
\author{V. M. Sauer}
% \author{B. Author}%
 \email{vmsauer@csun.edu}
\affiliation{ 
Department of Mechanical Engineering, California State University, Northridge, CA, USA
}%

% \author{C. Author}
 % \homepage{http://www.Second.institution.edu/~Charlie.Author.}
% \affiliation{%
% Second institution and/or address%\\This line break forced% with \\
% }%

\date{\today}% It is always \today, today,

\begin{abstract}
This work presents an analytical investigation of the hydrodynamic entrance region in axisymmetric laminar flows through slender converging pipes. Extending previous analyses for straight pipes, the model radially divides the flow into a viscous wall region and a central core where both inertia and viscous effects are important. The study analyzes the impact of the inlet Reynolds number and inlet angle on the developing velocity profile and pressure drop. Results show that a converging geometry, which imposes a favorable pressure gradient, significantly shortens the hydrodynamic entrance length compared to a straight pipe. Analytical solutions show good agreement with numerical simulations.
\end{abstract}

\maketitle

% \section{\label{sec:intro}Introduction}

The study of internal fluid flow in conduits with varying cross-sections is a classical problem of fundamental importance. Beyond its basic role in viscous–inertial interactions, it has practical relevance in various contexts, including propulsion systems, diffusers, nozzles, heat exchangers, and physiological transport. Although the entrance flow in a straight, uniform pipe is a well-known problem, the interaction between the developing boundary layer and the pressure gradient caused by a varying geometry introduces more complexity: in a converging section, a favorable pressure gradient accelerates the core flow. A central aspect of such internal flows is the hydrodynamic entrance region, where the velocity profile evolves from its initial inlet state to a fully developed, steady form.

The case of a straight circular pipe has been particularly well studied, forming the basis of classical entrance-region theory. Early work by \citet{langhaar_steady_1942} provided one of the first systematic descriptions of the ``transition length'' by linearizing the governing equations and deriving approximate relations for pressure loss. This approach was refined by \citet{campbell_flow_1963}, who emphasized the structure of the entrance profile, and by \citet{lundgren_pressure_1964}, who derived general expressions for the entrance-region pressure drop applicable to ducts of arbitrary cross-section. Further studies addressed specific geometries, including annular ducts \citep{sparrow_developing_1964} and rectangular sections \citep{sparrow_flow_1964,mccomas_hydrodynamic_1967}, and extended the framework to heat transfer problems \citep{kakac_analysis_1969}.

Asymptotic and analytical techniques also contributed significantly to the theoretical description of developing laminar flows. \citet{vandyke_entry_1970} obtained uniformly valid asymptotic solutions for channel entry flows at high Reynolds number, while \citet{fargie_developing_1971} and \citet{mohanty_laminar_1979} further clarified the division of the entrance into an inviscid core and viscous near-wall subregions. More recently, \citet{durst_development_2005} revisited the problem with careful experiments and simulations, improving hydrodynamic-entrance-length correlations. Most notably, \citet{kim_analytical_2024} introduced a new analytical solution to the parabolized Navier–Stokes equations for developing laminar pipe flows, which demonstrated that the approach to similarity is not monotonic but involves a near-wall velocity overshoot -- an effect overlooked in earlier models. This refinement provides a crucial starting point for further theoretical developments.

In contrast, the corresponding theory for ducts with slowly varying cross-sectional area remains less developed, despite their ubiquity. The classical Jeffery–Hamel solution \citep{jeffery_twodimensional_1915,hamel_spiralformige_1917} describes laminar flow in a two-dimensional wedge and illustrates the inherent tendency of diverging flows toward separation. However, its idealized geometry limits its applicability to two-dimensional channels. To address this, \citet{williams_viscous_1963} introduced a theoretical framework for incompressible viscous flow in slender pipes, where the radius varies slowly in the axial direction. This slender approximation, in which radial pressure gradients vanish at leading order, yields a parabolic system analogous to boundary-layer theory. Subsequent extensions \citep{blottner_numerical_1977,daniels_high_1979,eagles_steady_1982,kotorynski_viscous_1995} clarified the mathematical structure of the model and demonstrated its usefulness in describing both compressible and incompressible flows with mild axial area variation.

Several analytical and numerical investigations have attempted to capture the development of flows in converging and diverging channels. \citet{atabek_development_1973} provided an approximate analytical solution for converging geometries, while \citet{dennis_flow_1997} demonstrated that diverging channels can support multiple steady solutions, with branch selection depending sensitively on the inlet profile. \citet{garg_flow_1987} and \citet{mutama_investigation_1993} employed full numerical simulations to characterize developing flows in converging–diverging geometries, while \citet{gepner_flow_2016} investigated periodic converging–diverging channels, showing how repeated contractions and expansions reorganize the developing profile and enhance mixing. Diverging geometries have also been examined in the context of stability: \citet{sahu_stability_2005} demonstrated that even a small divergence introduces a finite critical Reynolds number, in contrast to the unconditional linear stability of straight-pipe flow; this prediction was later confirmed experimentally by \citet{peixinho_transition_2013}. More recent computational studies \citep{sparrow_flow_2009} mapped the onset of separation in diffusers, demonstrating that laminar flows at modest Reynolds numbers are far more sensitive to divergence angle than previously assumed.
\begin{figure}
    \centering
    \includegraphics{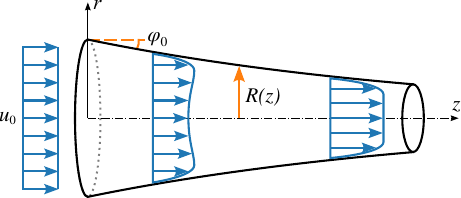}
    \caption{Schematic representation of a converging slender pipe.}
    \label{fig:schematic}
\end{figure}

Despite the extensive body of work on developing flows in straight pipes and quasi-developed flows in slender channels, to the author's knowledge, a comprehensive theoretical description of the hydrodynamic entrance region in slender converging pipes has yet to be developed. While \citet{williams_viscous_1963} established the governing equations for slender channels, and \citet{kim_analytical_2024} recently refined the entrance-region model for straight pipes to capture near-wall velocity overshoots, they do not provide a unified analytical description for the entrance region of converging pipes. Existing similarity solutions, such as the Jeffery-Hamel flow, only describe the fully developed asymptotic state and fail to capture the dynamic evolution of the velocity profile from the inlet. Additionally, classical boundary-layer integral methods for ducts with varying cross-sectional areas often overlook inertial effects that cause velocity overshoot. Although numerical methods can solve for these flows, a robust analytical model is essential for providing fundamental physical understanding and a rapid predictive tool. This work bridges that gap by extending the two-region model to variable-area pipes and providing a closed-form prediction of entrance length reduction in converging geometries.

Building on the two-region model of the entrance in straight pipes introduced by \citet{kim_analytical_2024} and embedding it within the slender-channel framework of \citet{williams_viscous_1963}, this work presents a complete analytical description of the developing laminar flow in pipes with mild convergence (\cref{fig:schematic}). The model divides the cross-section into a wall shear layer, dominated by viscous diffusion and pressure gradient, and an inertia-decaying core, where inertia and viscosity interact. Matching conditions at the interface ensure continuity of velocity and shear. This yields closed-form expressions for the velocity field, the pressure gradient, and the entrance length as functions of the Reynolds number and the wall slope. Comparisons with numerical simulations of laminar Navier–Stokes equations confirm the validity of the theory for converging pipes.

The manuscript begins with the theoretical formulation, covering the slender-pipe approximation and the two-region integral analysis. It then explains the numerical validation methods. Next, it presents the results, which include velocity profiles, pressure drop, and entrance length, before concluding with final remarks.

% \section{Theoretical Model}

A uniformly distributed flow with axial velocity $u_0$ enters the pipe from the left at $z = 0$, where $z$ is the axial coordinate, as shown in the schematic representation in \cref{fig:schematic}. The pipe geometry is defined by the varying radius $R(z)$. The assumption of a slender pipe implies that the radius varies slowly (gradually) relative to the axial length scale. The angle between the wall and the axial direction is represented by $\varphi$. The tangent of the inlet angle $\varphi_0$ (the wall angle $\varphi$ at $z=0$), $\tan{\varphi_0} = dR/dz|_0$, is considered a prescribed parameter.

The developing flow of a viscous, incompressible fluid in an axisymmetric slender pipe is governed by the steady conservation of mass and linear momentum equations. For flows with moderate to high inlet Reynolds numbers, $\Rey_0 \gg 1$, the conservation equations of mass and linear momentum in the axial and radial direction are given in non-dimensional form, respectively, as
\begin{subequations}
    \begin{gather}
        \label{eq:cons_mass}
        \frac{\partial u}{\partial z}
        + \frac{1}{r} \frac{\partial}{\partial r} (rv)
        = 0
        \\
        \label{eq:cons_mom_z}
        u \frac{\partial u}{\partial z}
        + v\frac{\partial u}{\partial r}
        = -\frac{\partial p}{\partial z}
        + \frac{2}{\Rey_0} \left[
            \frac{1}{r} \frac{\partial}{\partial r} \left(
                r \frac{\partial u}{\partial r}
            \right)
        \right]
        \\
        \label{eq:mom_cons_r}
        \frac{\partial p}{\partial r}=0
    \end{gather}
\end{subequations}
where $u$ and $v$ are the axial and radial velocity components scaled by the inlet average velocity $\hat{U}_0$. The hat decorator ($\hat{\phantom{a}}$) represents dimensional quantities. The pressure $p$ is scaled by the dynamic pressure $\hat{\rho} \hat{U}_0^2$, where $\hat{\rho}$ is the fluid density, and $r$ and $z$ are the radial and axial coordinates scaled by the inlet pipe radius $\hat{R}_0$. The flow Reynolds number is defined as $\Rey_0 = 2 \hat{R}_0 \hat{U}_0 / \hat{\nu}$, where $\hat{\nu}$ is the kinematic viscosity. \Cref{eq:cons_mass,eq:cons_mom_z,eq:mom_cons_r} are valid provided that the pipe length $L$ is large compared to the inlet radius, i.e., $L = \hat{L}/\hat{R}_0 \gg 1$.

To account for the slowly varying pipe radius, the coordinate transformation proposed by \citet{williams_viscous_1963} is employed,
\begin{equation}
    \xi = z
    \quad \text{and} \quad
    \eta = \frac{r}{R(\xi)}
\end{equation}
where $R(\xi)$ is the non-dimensional local pipe radius. Following \citet{kim_analytical_2024}, the flow is divided into two concentric regions: an inertia-decaying core ($0 \leqslant \eta \leqslant \eta_\delta$) and a wall shear layer ($\eta_\delta \leqslant \eta \leqslant 1$), with $\eta_\delta(\xi)$ marking their interface. In the wall layer, viscous forces dominate, whereas in the core, the flow inertia is influenced by the pressure gradient and shear force.

In the inertia-decaying core, the axial momentum equation, \cref{eq:cons_mom_z}, is approximated as
\begin{equation}\label{eq:cons_core}
    \frac{1}{2}\frac{\mathrm{d}u_0^2}{\mathrm{d}\xi}
    = -\frac{dp}{d\xi}
    + \frac{1}{R^2} \frac{2}{\Rey_0} \left[
        \frac{1}{\eta} \frac{\partial}{\partial\eta} \left(
            \eta\frac{\partial u_c}{\partial\eta}
        \right)
    \right]
\end{equation}
where $u_c$ is the axial velocity in the inertial-decaying core and $u_0(\xi)$ is the axial velocity at the centerline $(\eta = 0)$. In the wall shear layer, advection is neglected, and the momentum balance simplifies to
\begin{equation}\label{eq:cons_wall}
    0 =
    - \frac{dp}{d\xi}
    + \frac{1}{R^2} \frac{2}{\Rey_0} \left[
        \frac{1}{\eta} \frac{\partial}{\partial\eta} \left(
            \eta\frac{\partial u_w}{\partial\eta}
        \right)
    \right]
\end{equation}
where $u_w$ is the axial velocity in the wall region. The velocity field is subject to the no-slip condition at the wall, $u_w(1,\xi)=0$, and symmetry at the centerline, $\partial u_c / \partial \eta|_{\eta=0} = 0$. At the interface $\eta = \eta_\delta$, the velocity and its radial gradient are continuous:
\begin{equation}\label{eq:interface}
    u_c(\eta_\delta, \xi)
    = u_w(\eta_\delta, \xi)
    \quad \text{and} \quad
    \left. \frac{\partial u_c}{\partial\eta} \right|_{\eta_\delta}
    = \left. \frac{\partial u_w}{\partial\eta} \right|_{\eta_\delta}
\end{equation}

% \subsection{Analytical Solution}

The velocity profiles for the two regions are obtained from \cref{eq:cons_core,eq:cons_wall}. Integrating \cref{eq:cons_wall} for the wall shear layer and applying the no-slip condition yields
\begin{equation}
    u_w(\eta, \xi)
    = \frac{\Rey_0}{8} \frac{dp}{d\xi} R^2 (\eta^2 - 1) + a_1 \ln\eta
\end{equation}
where $a_1$ is an integration constant. Similarly, integrating \cref{eq:cons_core} for the inertia-decaying core and applying the centerline symmetry condition gives
\begin{equation}
    u_c(\eta, \xi)
    = \frac{\Rey_0}{8} R^2 \left(\frac{1}{2}\frac{\mathrm{d}u_0^2}{\mathrm{d}\xi} + \frac{dp}{d\xi}\right)\eta^2 + b_2
\end{equation}
where $b_2$ is an integration constant. Applying the interface conditions, \cref{eq:interface}, allows for the determination of $a_1$ and $b_2$.

Denoting
\begin{equation}\label{eq:Du_Dp}
    D_u
    = \Rey_0 \left(
        \frac{1}{2} \frac{d u_0^2}{d\xi}
    \right) R^4
    \quad \text{and} \quad
    D_p
    = \Rey_0
    \left(
        - \frac{d p}{d\xi}
    \right) R^4
\end{equation}
which represent dimensionless parameters related to inertial effects and axial pressure drop, the resulting velocity profiles are obtained as
\begin{gather}
    \label{eq:vel_core}
    u_c = u_0(1-\eta^2) + \frac{D_u}{8} \left[
        (1-\eta_\delta^2)
        + \eta_\delta^2 \ln\eta_\delta^2
    \right] \left(
        \frac{\eta}{R}
    \right)^2
    \\
    \label{eq:vel_wall}
    u_w = u_c + \frac{D_u}{8} \left[
        1
        - \left(
                \frac{\eta}{\eta_\delta}
            \right)^2
        + \ln\left(
            \frac{\eta}{\eta_\delta}
        \right)^2
    \right] \left(
        \frac{\eta_\delta}{R}
    \right)^2
\end{gather}
whereas the axial pressure drop is determined as
\begin{equation}\label{eq:pressure_drop}
    D_p =
    8 R^2 u_0
    + D_u \left(
        1
        - \ln{\eta_\delta^2}
    \right) \eta_\delta^2
\end{equation}

To solve for the unknowns $u_0(\xi)$ and $\eta(\xi)$, two governing ordinary differential equations (ODEs) are derived. Applying the global mass conservation constraint, $\int_{0}^{1}{u \eta d\eta} = 1/(2R^2)$, yields an algebraic relation for $D_u$,
\begin{equation}\label{eq:Du}
    D_u
    = \frac{
        8 \left(
            R^2 u_0 - 2
        \right)
    }{
        \eta_\delta^2 \left(
            1 - \eta_\delta^2 + \ln{\eta_\delta^2}
        \right)
    }
\end{equation}
The Kármán-Pohlhausen momentum integral technique is applied to the governing equations, resulting in an equation for the axial evolution of the global momentum, $\Theta(\xi) = R^2 \int_{0}^{1}{u^2 \eta d\eta}$,
\begin{equation}
    \frac{d\Theta}{d\xi}
    = \frac{
        D_u \eta_\delta^2
    }{
        2 \Rey_0 R^2
    }
\end{equation}

The integral for $\Theta(\xi)$ is evaluated using the velocity profiles from \cref{eq:vel_core,eq:vel_wall}, yielding a complex algebraic expression
\begin{multline}\label{eq:theta_eta}
    \Theta(\xi)
    = 
    R^2 \bigg\{
        \frac{u_0^2}{6} +
        \\
        \frac{1}{12} \bigg[
            \frac{
                5
                - \eta_\delta^2 (4 + \eta_\delta^2)
                + (5 + 2 \ln{\eta_\delta^2} + \eta_\delta^4) \ln{\eta_\delta^2}
            }{
                \left(
                    1
                    + \ln{\eta_\delta^2}
                    - \eta_\delta^2
                \right)^2
            } \left(
                u_0 - \frac{2}{R^2}
            \right)^2
            \\
            - \frac{
                    5
                    + \eta_\delta^2 (\eta_\delta^2 - 6)
                    + 4 \ln{\eta_\delta^2}
            }{
                1
                + \ln{\eta_\delta^2}
                - \eta_\delta^2
            } \left(
                u_0 - \frac{2}{R^2}
            \right) u_0
            % } \left(
            %     u_0 - \frac{2}{R^2}
            % \right) u_0
            % + 2 u_0^2
        \bigg]
    \bigg\}
\end{multline}
which, in the case of straight pipes ($R=1$), attains a fixed value of 2/3 for $\xi \rightarrow \infty$. In this limit, the terms in the square brackets become zero, since~\citep{kim_analytical_2024} $u_0 = 2$ at $\xi \gg 1$.

The system is closed by combining \cref{eq:Du} with the definition of $D_u$, \cref{eq:Du_Dp}, which provides an ODE for $u_0$. The resulting system of first-order ODEs for $u_0$ and $\Theta$ is solved numerically in Python using the SciPy wrapper to the LSODA Fortran solver from ODEPACK \citep{hindmarsh_odepack_1983}, where $\eta_\delta$ is determined iteratively from \cref{eq:theta_eta} [using SciPy's implementation of Brent’s root finding method \citep{brent_algorithms_2013}] at each axial step. The system is integrated starting from $\xi=0$ with initial conditions corresponding to a uniform inlet profile, namely $u_0(0)=1$ [since $R(0)=1$] and $\Theta(0)=1/2$, which is obtained from the global momentum integral.

\begin{figure}
    \centering
    \subfloat{
         \includegraphics{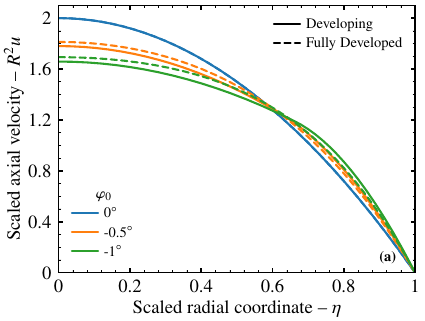}
         \label{fig:fully_u_profile}
    }
    \hspace{0.5em}
    \subfloat{
         \includegraphics{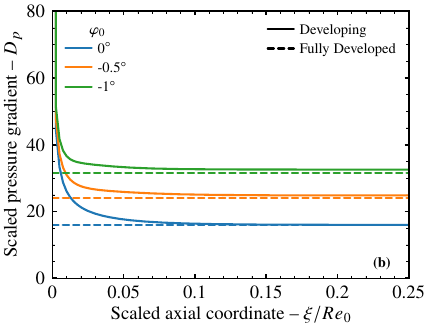}
         \label{fig:fully_dp}
    }
    \caption{Comparison between developing (solid lines) and fully developed (dashed lines) solutions at selected inlet angles for $\Rey_0 = \num{250}$. (a)~Radial scaled axial velocity profiles, and (b)~Axial scaled pressure gradient profiles.}
    \label{fig:fully_comparison}
\end{figure}

% \section{Results and Dicussion}

The analytical model is evaluated for Reynolds numbers $\Rey_0 = \num{250}$ and \num{500}. This model applies to pipes with a gradual change in radius along the length, but we will focus on results for conditions in which the axial velocity becomes self-similar after the flow development region, specifically to determine the entrance length. As a result, the pipe geometry is defined by the tangent of the inlet angle $\varphi_0$ (or $dR/d\xi|_0$), with values between \qty{-2}{\degree} (\num{-0.03492}) and \qty{0}{\degree} (straight pipe) being considered.

Results are presented to first establish the theoretical consistency of the model in the fully developed limit, followed by an analysis of the developing flow characteristics and a validation against numerical simulations.

\begin{table*}
    \setlength{\tabcolsep}{0.6em}
    \centering
    \begin{tabular}{c S[table-format=+1.2] S[table-format=+1.2] S[table-format=1.4] S[table-format=2.2] S[table-format=2.2] S[table-format=1.2] S[table-format=1.2] S[table-format=1.2] S[table-format=1.2]}
        \toprule
         & & & & \multicolumn{3}{c}{Pressure Gradient} & \multicolumn{3}{c}{Centerline Velocity} \\
        \cmidrule(lr){5-7} \cmidrule(lr){8-10}
        {$Re_0$} & {$\varphi_0$} & {$B$} & {$\ell_e / \Rey_0$} & {$\alpha$} & {$D_p$} & {\% Diff.} & {$F_0$} & {$R^2 u_0$} & {\% Diff.}\\
        \midrule
        Laminar & \hspace{0.8em}\SI{0.0}{\degree} & 0.00 & 0.1174 & 16.00 & 16.00 & 0.03 & 2.00 & 2.00 & 0.02 \\
        \addlinespace % \cmidrule{1-5}
        \multirow{4}{*}{250}
        & \SI{-0.5}{\degree} & -2.18 & 0.1027 & 24.13 & 24.80 & 2.81 & 1.81 & 1.78 & 1.81 \\
        & \SI{-1.0}{\degree} & -4.36 & 0.0764 & 31.47 & 32.55 & 3.43 & 1.69 & 1.66 & 2.07 \\
        & \SI{-1.5}{\degree} & -6.55 & 0.0610 & 38.36 & 39.74 & 3.59 & 1.61 & 1.58 & 1.91 \\
        & \SI{-2.0}{\degree} & -8.73 & 0.0502 & 44.96 & 46.58 & 3.60 & 1.55 & 1.52 & 1.63 \\
        \addlinespace % \cmidrule{1-5}
        \multirow{4}{*}{500}
        & \SI{-0.5}{\degree} & -4.36 & 0.0764 & 31.47 & 32.55 & 3.43 & 1.69 & 1.66 & 2.07 \\
        & \SI{-1.0}{\degree} & -8.73 & 0.0502 & 44.95 & 46.57 & 3.60 & 1.55 & 1.52 & 1.63 \\
        & \SI{-1.5}{\degree} & -13.09 & 0.0364 & 57.53 & 59.55 & 3.51 & 1.46 & 1.44 & 1.06 \\
        & \SI{-2.0}{\degree} & -17.46 & 0.0289 & 69.59 & 71.94 & 3.37 & 1.40 & 1.39 & 0.60 \\
        \bottomrule
    \end{tabular}
    \caption{Comparison of developing flow model parameters with the fully developed similarity solution. Functions $D_p$ and $R^2 u_0$ evaluated at $\xi/\Rey_0 =0.25$.}
    \label{tab:comparison}
\end{table*}

% \subsection{Asymptotic Behaviour: Fully Developed Flow}

To validate the current model, we investigate its asymptotic behavior by comparing the solution in the far-downstream region with the classical similarity solution for fully developed flow in slender pipes. \citet{williams_viscous_1963} identified such fully developed flows as the only cases where similarity solutions can be applied in slender pipes. Therefore, we limit our analysis to converging pipes that exhibit a specific cross-sectional variation along the axial direction. If we were to analyze other configurations, the flow field would not achieve self-similarity.

For a fully developed flow, the velocity components can be expressed in terms of a similarity function $F(\eta)$ as $u=F/R^2$ and $v = (dR/d\xi) \eta F/R^2$. Substituting these expressions into the momentum equation yields the governing ordinary differential equation \citep{sauer_laminar_2024}
\begin{equation}\label{eq:developed}
    F'' + \frac{1}{\eta} F' + B F^2 = - \frac{\alpha}{2}
\end{equation}
where $B = \Rey_0 (1/R)(dR/d\xi)$ and $\alpha = Re_0 (-dp/d\xi) R^4$ are constants, requiring that
\begin{equation}
    R(\xi) = \exp{(B \xi / \Rey_0)}
\end{equation}

For a given $B$, the pressure gradient parameter $\alpha$ is an eigenvalue of the problem. The solution of \cref{eq:developed} is determined considering the boundary conditions $F(1)=0$, $F'(0)=0$, and $F(0) = F_0$. The similarity function $F(\eta)$ is obtained iteratively by adjusting the parameter $F_0$ to satisfy the condition $\int_0^1{F \eta d\eta} = 1/2$.

As $\xi \rightarrow \infty$, the developing flow must converge to a self-similar state. This implies that the scaled axial velocity profile from the present model, $R(\xi)^2 u(\eta,\xi)$, should approach the similarity function $F(\eta)$. Concurrently, the dimensionless pressure gradient parameter from the developing model, $D_p(\xi)$, should converge to the constant eigenvalue $\alpha$ of the fully developed problem.

\Cref{fig:fully_u_profile} compares the asymptotic velocity profile from the developing flow model with the fully developed similarity solution for a converging pipe at $\Rey_0 = \num{250}$ for $\varphi_0 = \qty{0}{\degree}, \qty{-0.5}{\degree}$, and \qty{-1}{\degree}. The dashed lines represent the fully developed solution $F(\eta)$, whereas the solid lines show the developing flow solution $R^2 u$ evaluated at a large axial distance ($\xi/\Rey_0 = 0.25$). A small deviation between the profiles is observed near $\eta = 0.5$. This is inherent to the two-region approximation, which imposes continuity of the function and its first derivative but does not guarantee continuity of higher-order derivatives at the interface $\eta_\delta$. However, the general agreement confirms that the developing flow solution converges to the fully developed profile.

The convergence of the pressure gradient parameter is shown in \cref{fig:fully_dp} for $Re_0 = 250$. The value of $D_p$ from the developing model is plotted against the axial coordinate $\xi/\Rey_0$. Horizontal (asymptotic) dashed lines representative of the pressure drop $\alpha$ for fully developed flow are also included. It is observed that $D_p$ rapidly approaches the fully developed solution for $\varphi_0 = \qty{0}{\degree}, \qty{-0.5}{\degree}$, and \qty{-1}{\degree}.

A quantitative comparison of the asymptotic pressure gradient parameter is presented in \cref{tab:comparison} for $Re_0$ of \num{250} and \num{500} at several inlet angles $\varphi_0$. The table lists the eigenvalue $\alpha$ obtained from the fully developed model and the asymptotic value of $D_p$ (at $\xi/\Rey_0 = 0.25$) from the present developing flow model. The relative difference is less than \qty{4}{\percent} for all cases. The scaled centerline velocity values for the fully developed ($F_0$) and developing ($R^2 u_0$) solutions are also shown. The relative difference of \qty{2}{\percent} or less demonstrates good agreement between solutions, confirming the consistency of the present model with the established theory for fully developed slender pipe flows.

\begin{figure}
    \centering
    \subfloat{
         \includegraphics{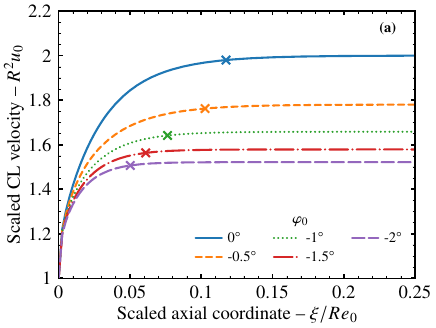}
         \label{fig:developing_conv}
    }
    \hspace{0.5em}
    \subfloat{
         \includegraphics{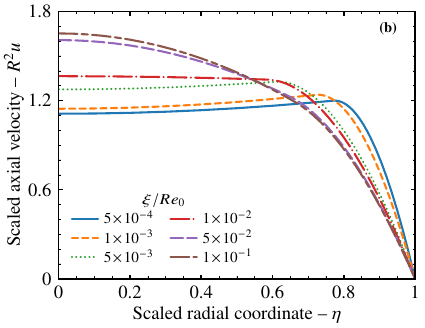}
         \label{fig:developing_ucl}
    }
    \hspace{0.5em}
    \subfloat{
         \includegraphics{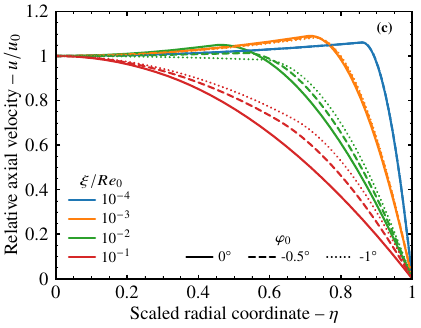}
         \label{fig:developing_overshoot}
    }
    \caption{Axial velocity profiles for $\Rey_0 = \num{250}$. (a)~Centerline profiles at selected inlet angles, (b)~Profiles at selected axial positions for $\varphi_0 = \SI{-1}{\degree}$, and (c)~Relative axial velocity profiles at selected axial positions and inlet angles. Markers $\times$ denote the entrance length $\ell_e/Re_0$ for each case.}
    \label{fig:developing_slender}
\end{figure}

% \subsection{Developing Flow Characteristics\label{sec:developing}}

Upon confirmation that the solution derived from the analytical model accurately represents the fully developed region, the focus shifts to analyzing the developing flow.

\Cref{fig:developing_conv} shows the evolution of the centerline velocity, $u_0$, along the non-dimensional axial coordinate, $\xi/\Rey_0$, for selected geometries at $\Rey_0 = \num{250}$. In all cases, the centerline velocity accelerates from its initial value and approaches an asymptotic, fully developed value. For the straight pipe, $u_0$ approaches the classical value of \num{2.0}. The converging geometries, with their favorable pressure gradient, cause a more rapid acceleration and a shorter hydrodynamic entrance length, $\ell_e$, which is assumed as the axial location where~\citep{kim_analytical_2024} $R^2 u_0(\ell_e) = 0.99 F_0$. As the convergence angle increases ($B \ll -1$), the effect of the strong favorable pressure gradient becomes dominant, leading to the expected rapid decrease in entrance length. The following polynomial fitting of degree two is obtained for the interval $\num{-17.5} \leqslant B \leqslant \num{0}$, considering the values from \cref{tab:comparison},
\begin{equation}\label{eq:entrance_length}
    \frac{\ell_e}{\Rey_0} = 
        \num{3.045e-4} B^2 + \num{1.033e-2} B + 0.1174 %, \quad -17.5 \leqslant B \leqslant 0
\end{equation}
which can be used to predict $\ell_e$. The root mean square error fitting is about \qty{0.1}{\percent}.

It is worth mentioning that the specific coefficients in \cref{eq:entrance_length} depend on the chosen ratio $\lambda$ (ratio of the central axis velocity to the fully developed value, $\lambda = R^2 u_0(\ell_e)/F_0$), where $\lambda = 0.99$ is the conventional definition of the entrance length. Selecting a stricter ratio would increase the constant term but would not alter significantly the functional dependence on $B$. Calculated values of $\ell_e / \Rey_0$ and $R^2 u_0(\ell_e)$ for selected $\lambda$ values are shown in \cref{tab:other} for reference. The values of $B$ are chosen based on \cref{tab:comparison}.

\begin{table}
    \setlength{\tabcolsep}{0.6em}
    \centering
    \begin{tabular}{S[table-format=+2.2] S[table-format=1.4] S[table-format=1.2] S[table-format=1.4] S[table-format=1.2] S[table-format=1.4] S[table-format=1.2]}
        \toprule
         & \multicolumn{2}{c}{$\lambda = 0.98$} & \multicolumn{2}{c}{$\lambda = 0.99$} & \multicolumn{2}{c}{$\lambda = 0.999$} \\
        \cmidrule(lr) {2-3} \cmidrule(lr){4-5} \cmidrule(lr){6-7}
        {$B$} & {$\ell_e / \Rey_0$} & {$R^2 u_0$} & {$\ell_e / \Rey_0$} & {$R^2 u_0$} & {$\ell_e / \Rey_0$} & {$R^2 u_0$} \\
        \midrule
        \addlinespace % \cmidrule{1-5}
          0.00 & 0.0942 & 1.96 & 0.1174 & 1.98 & 0.1957 & 2.00 \\
         -2.18 & 0.0809 & 1.74 & 0.1027 & 1.76 & 0.1754 & 1.78 \\
         -4.36 & 0.0601 & 1.62 & 0.0764 & 1.64 & 0.1288 & 1.66 \\
         -6.55 & 0.0476 & 1.55 & 0.0610 & 1.56 & 0.1095 & 1.58 \\
         -8.73 & 0.0390 & 1.49 & 0.0502 & 1.51 & 0.0894 & 1.52 \\
        -13.09 & 0.0282 & 1.42 & 0.0364 & 1.43 & 0.0623 & 1.44 \\
        -17.46 & 0.0221 & 1.37 & 0.0289 & 1.38 & 0.0515 & 1.39 \\
        \bottomrule
    \end{tabular}
    \caption{Comparison of developing flow model entrance length and fully developed velocity for different ratios $\lambda$.}
    \label{tab:other}
\end{table}

The development of the axial velocity profile, $u$, is shown in \cref{fig:developing_ucl} for the converging case with $\Rey_0 = \num{250}$ and an inlet angle of \qty{-1}{\degree} at selected axial locations. Near the inlet ($\xi/\Rey_0 \ll 1$), the velocity profile is relatively flat in the core. As the flow moves downstream, the boundary layer grows, and the profile gradually evolves toward its fully developed shape. The resulting profile is less parabolic (more plug-like) than the Poiseuille profile. This is a direct consequence of the favorable pressure gradient, which accelerates the core flow relative to the fluid near the wall, creating a more uniform velocity distribution across the central region (cf.~\cref{tab:comparison}).

A velocity overshoot occurs near the wall before the flow becomes fully developed, as shown in \cref{fig:developing_ucl}. To examine how the overshoot magnitude qualitatively depends on inlet angle, a comparison between the axial velocity profile relative to the centerline velocity is presented in \cref{fig:developing_overshoot}. Near the inlet, the curvature of the streamlines is more pronounced due to the rapid transition from a flat to a developing profile~\cite{kim_analytical_2024}, which causes nonzero radial pressure gradients. This behavior is not captured by the simplified conservation equations in the model, \cref{eq:mom_cons_r}, resulting in similar axial velocity profiles for $\xi/\Rey_0 < \num{e-2}$. However, as the flow proceeds downstream, larger convergence angles lead to more pronounced core flow acceleration due to geometric confinement. Although the absolute velocity increases everywhere, the relative overshoot is suppressed as the favorable pressure gradient stabilizes the boundary layer. The analytical model captures this non-monotonic behavior, predicting the approximate location and magnitude of the peak velocity within the boundary layer.

% \subsection{Numerical Validation}

To validate the analytical model, solutions for selected cases are compared with numerical simulations obtained using the commercial software ANSYS\textsuperscript{\textregistered} Fluent 2024 R1. The numerical analysis solves the full laminar Navier-Stokes equations, including axial diffusion and radial pressure gradients, in a three-dimensional domain. The mesh is generated in ANSYS\textsuperscript{\textregistered} SpaceClaim using an O-type structured grid for the cross-section, with refinement at the pipe inlet and walls to account for flow development and boundary-layer effects with a bias factor of 4, resulting in an average of \num{10e6} hexahedral cells. The Coupled algorithm is used for pressure-velocity coupling, whereas a second-order upwind method is selected for the momentum fluxes. A uniform axial velocity profile is specified at the inlet ($z = 0$) to simulate the developing flow condition, with no-slip stationary wall and pressure-outlet boundary conditions applied to their respective surfaces. A grid independence study was performed, confirming that the mesh resolution is sufficient to ensure results free from significant discretization error. Solutions are obtained by satisfying the convergence criterion with residuals of \num{e-5} for all equations.

\begin{figure}
    \centering
    \subfloat{
        \includegraphics{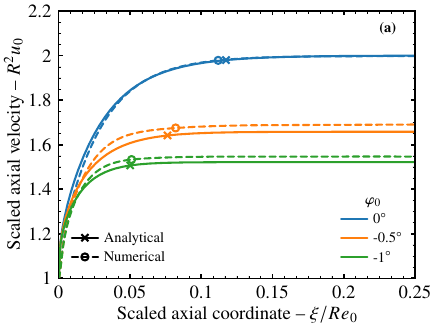}
        \label{fig:numerical_u}
    }
    \hspace{0.5em}
    \subfloat{
        \includegraphics{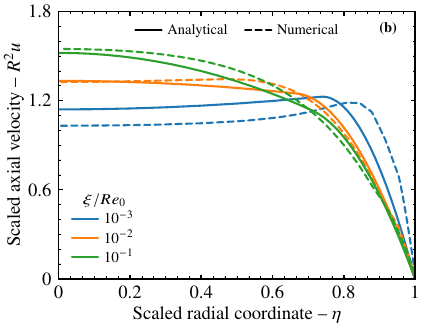}
        \label{fig:numerical_p}
    }
    \caption{Analytical (solid lines) and numerical (dashed lines) axial velocity solutions for $\Rey_0 = \num{500}$. (a)~Variation in the axial direction for selected $\varphi_0$ (markers represent entrance lengths), and (b)~Profiles at selected axial positions for $\varphi_0 = \qty{-1}{\degree}$.}
    \label{fig:numerical}
\end{figure}

\Cref{fig:numerical} compares the pressure drop predicted by the analytical model with numerical results for converging geometries at $\Rey_0 = \num{500}$ and $\varphi_0 = \qty{0}{\degree}$, \qty{-0.5}{\degree}, and \qty{-1}{\degree}. \Cref{fig:numerical_u} shows that the centerline velocity $u_0$ reaches a value of \num{2.0} in the fully developed region of straight pipes, consistent with classical theory. The converging pipe exhibits a much steeper acceleration, aligning with the analytical results. The numerical solution at $\xi/\Rey_0 = \num{0.25}$ nearly matches the fully developed solution from \cref{tab:comparison}. Furthermore, a slight deviation is observed between the numerical and analytical entrance lengths. \Cref{fig:numerical_p} displays the numerical and analytical solutions of scaled axial velocity profiles at selected axial locations for $\Rey = 500$ and $\varphi_0 = \qty{-1}{\degree}$. A larger deviation between the numerical and analytical profiles appears near the inlet ($\xi/Re_0 < \num{e-2}$), which is also seen in straight pipe solutions \citep{kim_analytical_2024}. This discrepancy results from the inherent limitations of the boundary-layer approximation near the inlet. Unlike the analytical model, the numerical solution of the Navier-Stokes equations captures the influence of radial pressure gradients that the slender approximation, \cref{eq:mom_cons_r}, does not. Nevertheless, the solution rapidly converges to the numerical data slightly downstream, confirming the robustness of the model over most of the entrance length.

The reduction in hydrodynamic entrance length as the convergence angle increases affects engineering design. In microfluidic nozzles and dispensing systems, a shorter entrance length allows flow to stabilize more quickly, enabling more compact designs while maintaining predictable flow. Accurate prediction of entrance length is crucial for estimating pressure drops in short converging sections, such as catheter tips or extrusion dies. The velocity overshoot in the developing region suggests that shear stress in converging vessels may exceed that predicted by fully developed models.

% \section{Conclusion}

An analytical model for developing laminar flow in slender converging pipes has been developed. The model extends the classical two-region analysis of \citet{kim_analytical_2024} to pipes with slowly varying cross-sections by incorporating the slender-pipe approximation of \citet{williams_viscous_1963}. The solution provides a complete description of the velocity field and pressure distribution throughout the hydrodynamic entrance region.

The model accurately predicts the evolution of the velocity profile from a uniform inlet condition to a fully developed state. The analytical solutions for centerline velocity, velocity profiles, and pressure drop exhibit good agreement with numerical simulations of the full Navier-Stokes equations for inlet Reynolds numbers of \num{250} and \num{500}, and inlet angles ranging from \qty{-2}{\degree} to \qty{0}{\degree}. The pipe geometry has a profound effect on the hydrodynamic entrance length. A converging geometry creates a favorable pressure gradient that accelerates flow development, resulting in a shorter entrance length compared to a straight pipe. The solution for the developing flow is shown to asymptotically converge to the classical similarity solution for fully developed slender pipe flow in the far-downstream limit, confirming the theoretical consistency of the model. Limitations of this work include the restriction to axisymmetric laminar flows; turbulent transition and three-dimensional effects (e.g., non-circular conduits) remain areas for future investigation.

The theoretical framework presented herein provides fundamental physical insight into the interplay between viscous boundary layer growth and pressure gradients imposed by varying geometries. Furthermore, it serves as a robust, computationally efficient predictive tool for the design and analysis of systems involving flows in slender, non-uniform conduits.

\begin{acknowledgments}
This work was supported by the National Science Foundation through grant \# 2302003.
\end{acknowledgments}

\section*{AUTHOR DECLARATIONS}

The authors report no conflict of interest.

\section*{Data Availability Statement}

The data that support the findings of this study are available from the corresponding author upon reasonable request.

\section*{References}
% \nocite{*}
\bibliography{references}% Produces the bibliography via BibTeX.

\end{document}